\documentclass{article}
\usepackage{spconf,amsmath,graphicx}
\usepackage{booktabs,adjustbox,makecell}

\usepackage[labelformat=simple]{subcaption}

\usepackage[font=small,labelfont=bf]{caption}
\usepackage{multirow}
\usepackage{adjustbox}
\usepackage{flushend}

\title{MULTI-CLASS SPECTRAL CLUSTERING WITH OVERLAPS FOR\\SPEAKER DIARIZATION}
\name{Desh Raj$^1$, Zili Huang$^1$, Sanjeev Khudanpur$^{1,2}$}
\address{
  $^1$Center for Language and Speech Processing \& $^2$Human Language Technology Center of Excellence \\ 
  The Johns Hopkins University, Baltimore, MD 21218, USA.}
\email{draj@cs.jhu.edu, \{hzili1,khudanpur\}@jhu.edu}
\vspace{-1em}
\begin{document}
%
\maketitle
\begin{abstract}
This paper describes a method for overlap-aware speaker diarization. Given an overlap detector and a speaker embedding extractor, our method performs spectral clustering of segments informed by the output of the overlap detector. This is achieved by transforming the discrete clustering problem into a convex optimization problem which is solved by eigen-decomposition. Thereafter, we discretize the solution by alternatively using singular value decomposition and a modified version of non-maximal suppression which is constrained by the output of the overlap detector. Furthermore, we detail an HMM-DNN based overlap detector which performs frame-level classification and enforces duration constraints through HMM state transitions. Our method achieves a test diarization error rate (DER) of 24.0\% on the mixed-headset setting of the AMI meeting corpus, which is a relative improvement of 15.2\% over a strong agglomerative hierarchical clustering baseline, and compares favorably with other overlap-aware diarization methods. Further analysis on the LibriCSS data demonstrates the effectiveness of the proposed method in high overlap conditions.
\end{abstract}
\section{Introduction}
\label{sec:intro}

Speaker diarization (or ``who spoke when?'') refers to the task of segmenting speech into homogeneous speaker-specific regions~\cite{Mir2012SpeakerDA, Tranter2006AnOO}. Conventional diarization systems~\cite{GarciaRomero2017SpeakerDU, Sun2018SpeakerDW} consist of four major components. First, a speech activity detection module removes the non-speech segments. Next, the speech regions of the recording are divided into small (often overlapping) segments and a pretrained speaker embedding extractor is used to obtain fixed-dimensional embeddings, such as i-vectors, or neural embeddings~\cite{Dehak2011FrontEndFA,Variani2014DeepNN,Snyder2018XVectorsRD}, for each segment. The embeddings are scored pairwise using a cosine or probabilistic linear discriminant analysis (PLDA) similarity metric, and clustering (agglomerative or spectral) is performed on the resulting affinity matrix until a stopping criterion is reached or until the desired number of speaker clusters is obtained. Finally, a resegmentation module~\cite{Sell2015DiarizationRI} may be used for frame-level refinement of the clustering output.

Although this approach has proved to be effective through use of deep neural network based speaker embeddings, it does not handle overlapping speaker segments, since the clustering process assigns each segment to exactly one speaker. Existing approaches to solve the overlap problem fall into two categories. In the first framework, an externally trained overlap detection module identifies frames in the recording which contain overlapping speech. This ``overlap detection'' may be performed using hidden Markov models (HMMs)~\cite{boakye2008overlapped, Huijbregts2009SpeechOD, Yella2012SpeakerDO} or neural networks~\cite{Geiger2013DetectingOS, Andrei2017DetectingOS, Hagerer2017EnhancingLR, Kunesov2019DetectionOO}. Once overlaps are detected, an ``overlap assignment'' stage assigns additional speaker labels to the overlapping frames. Recently, \cite{Bullock2019OverlapawareDR} proposed overlap-aware resegmentation, which leverages the variational Bayes (VB)-HMM method used originally for diarization in~\cite{Dez2018SpeakerDB}, and applied to resegmentation in~\cite{Sell2015DiarizationRI}. In the second framework, end-to-end systems~\cite{Fujita2020EndtoEndND,Huang2020SpeakerDW} are used to perform overlapping diarization in a supervised setting. 

In this paper, we focus on the former approach for overlap-aware speaker diarization. Specifically, we train an external overlap detector, and use its classification decision during clustering of the segment-level embeddings. Our method relies on the two-step clustering formulation proposed in~\cite{Yu2003MulticlassSC}. In the first step, the NP-hard discrete clustering problem is relaxed into a continuous version by ignoring the discrete constraints on the solution. The continuous problem thus obtained has a solution set generated through orthonormal transformations of eigenvectors of the normalized Laplacian. The second step involves ``optimal discretization'', which finds a discrete solution under the constraint that it is close (in Frobenius norm) to any of the relaxed solutions from the solution set obtained previously. We introduce overlap awareness in the discretization stage by modifying the ``sum-to-one'' constraint in this subproblem. This modification makes it possible to perform overlap-aware spectral clustering at no extra computational cost beyond computing the overlap decisions. Furthermore, we use the recently proposed $p$-binarization and normalized maximum eigengap (NME) techniques~\cite{Park2020AutoTuningSC} to self-tune the clustering process, thus requiring no hyperparameter tuning for estimating the number of speakers.

The remainder of this paper is organized as follows. We start by giving a detailed description of our method in Section~\ref{sec:method}, where we discuss $p$-binarization and NME for estimating the number of speakers followed by the mathematical formulation of multi-class spectral clustering. We then introduce our modification of the method to perform overlap-aware diarization. We describe our HMM-DNN based overlap detector in Section~\ref{sec:overlap}. This is followed by a description of our experimental setup and results in Sections~\ref{sec:experiments} and \ref{sec:results}, respectively. We present results on the AMI meeting corpus and the LibriCSS dataset, with detailed analysis on the performance of the method on different overlap conditions. In Section~\ref{sec:related}, we summarize previous work on spectral clustering for speaker diarization. We conclude with a discussion of future work in Section~\ref{sec:conclusion}. In the interest of reproducible research, we discuss our implementation in detail in Section~\ref{sec:implementation}, and our code has been made publicly available at: {\footnotesize \texttt{https://desh2608.github.io/pages/overlap-aware-sc}}.

\section{Methodology}
\label{sec:method}

Our diarization follows the conventional clustering method studied extensively in previous work~\cite{GarciaRomero2017SpeakerDU, Sun2018SpeakerDW}. We first obtain speech regions from the recording using an oracle speech activity detector (although any SAD can be used for this purpose). These are divided into small (overlapping) segments using a sliding window method (we used 1.5s segments with a stride of 0.75s in our experiments), and embeddings are extracted for each segment using an x-vector extractor~\cite{Snyder2018XVectorsRD}. Subsequently, these embeddings are clustered to obtain speaker groupings, and each cluster is labeled as a speaker.

We do not present the details of the x-vector extraction here; readers unfamiliar with how they are trained are referred to \cite{Snyder2018XVectorsRD}. Our focus is on the final clustering stage, and particularly on how to make the clustering process overlap-aware. In particular, given a sequence of segment embeddings $U = (\mathbf{u}_1, \ldots, \mathbf{u}_i, \ldots, \mathbf{u}_N)$, the objective is to compute a label sequence $L = (\boldsymbol{\ell}_1, \ldots, \boldsymbol{\ell}_i, \ldots, \boldsymbol{\ell}_N)$, where $\boldsymbol{\ell}_i$ may be a single label or, in case of overlapping segment, a tuple of multiple labels. For convenience, we assume that overlaps can occur between at most two speakers, so $\boldsymbol{\ell}_i$ will be a 2-tuple for an overlapping segment. Finally, suppose we have an overlap detector which decides, for each segment, whether or not it contains overlapping speech. We denote this as
\begin{equation}
\label{eqn:overlap}
f(U) = \mathbf{v}_{OL},
\end{equation}
where $\mathbf{v}_{OL} \in \{0,1\}^N$, and $\mathbf{v}_{OL}^i = 1$ indicates that ${\ell}_i$ is an $n$-tuple with $n>1$. Given $U$ and $\mathbf{v}_{OL}$, overlap-aware diarization seeks to compute an optimal label sequence $L$ which minimizes the diarization error. Since we do not additionally have information about the number of speakers $K$ in the recording, we first estimate it using the heuristic described in~\cite{Park2020AutoTuningSC}. Subsequently, we perform multi-class spectral clustering to group the $N$ segments into the estimated $\widehat{K}$ clusters using the optimal discretization procedure proposed in~\cite{Yu2003MulticlassSC}, where we make a key modification to constrain the optimization process on the output $\mathbf{v}_{OL}$ of our overlap detector.

\subsection{Estimating number of speakers}

We first use the method described in~\cite{Park2020AutoTuningSC} to estimate the number of speakers $\widehat{K}$ in the recording using an eigengap heuristic. Here, we describe this method briefly. 

Given $U$, we compute the affinity matrix $\mathbf{A} \in [-1,1]^{N\times N}$ of raw cosine similarity values. Then, $p$-binarization is performed on this matrix by replacing the $p$ highest similarity values in each row with 1, and the rest with 0, followed by a symmetrization operation, 
\begin{equation}
\bar{\mathbf{A}}_p = \frac{1}{2}(\mathbf{A}_p + \mathbf{A}_p^T).
\end{equation}
We compute the unnormalized Laplacian for this matrix,
\begin{equation}
\mathbf{L}_p = \mathbf{D}_p - \bar{\mathbf{A}}_p,
\end{equation}
where $\mathbf{D}_p = \mathrm{diag}\{d_1,\ldots,d_N\}$, $d_i = \sum_{n=1}^N a_{in}$, also known as the ``degree'' of node $i$.

The properties of the unnormalized Laplacian of the affinity matrix have been studied extensively~\cite{Luxburg2007ATO}, and it is known that $\mathbf{L}_p$ has $N$ non-negative, real eigenvalues $0=\lambda_1 \leq \lambda_2 \leq \ldots \leq \lambda_N$. Furthermore, an implication of the Davis-Kahan perturbation theory~\cite{Stewart1990MatrixPT} proposes an eigengap heuristic for the optimal number of clusters. In~\cite{Park2020AutoTuningSC}, the authors used this heuristic to estimate the optimal $p$ value for the binarization (described earlier). Specifically, let $\mathbf{e}_p$ denote the vector of differences in consecutive eigenvalues (in increasing order). We compute the quantities
\begin{equation}
g_p = \frac{\max(\mathbf{e}_p)}{\lambda_{p,N}+\epsilon},\quad \text{and} \quad r(p) = \frac{p}{g_p}.
\end{equation}
Then, the optimal $p$, i.e. $\hat{p}$ is the one that minimizes $r(p)$, and subsequently, the optimal number of clusters is given as
\begin{equation}
    \widehat{K} = \arg\max(\mathbf{e}_{\hat{p}}).
\end{equation}
In the multi-class spectral clustering procedure below, we will use this estimate $\hat{K}$ for the number of clusters, and drop the subscript $p$ from the matrices like $\mathbf{L}_p$, $\mathbf{D}_p$ and $\bar{\mathbf{A}}_p$ for brevity.

\subsection{Multi-class spectral clustering}

Bipartite graph partitioning using the affinity matrix Laplacian $\mathbf{L}$ is solved by node assignment based on the underlying Fiedler vector (eigenvector corresponding to the second smallest eigenvalue)~\cite{Fiedler1973AlgebraicCO}. The Ng-Jordan-Weiss algorithm~\cite{Ng2001OnSC} is a popular extension of this principle for multi-way partitioning of the graph. It applies K-means clustering on the first $K$ eigenvectors of $L$, i.e., in the $K$-eigenspace of the Laplacian. It is known that if the original samples are separable into $K$ groups using some transformation, then their projection on the $K$-eigenspace can be easily grouped using K-means clustering. Recent work on speaker diarization through spectral clustering of x-vectors, as in~\cite{Park2020AutoTuningSC} and ~\cite{Medennikov2020TargetSpeakerVA}, has employed this algorithm. However, there are two major limitations of this approach. First, the K-means clustering process may get stuck in bad local optima, particularly when the affinity matrix is noisy. Second, and particularly relevant for our case, it is difficult to extend this method to handle overlaps. To remedy these issues, we use an alternative formulation of spectral clustering, proposed in~\cite{Yu2003MulticlassSC}. 

Given $\mathbf{A}$ and $\mathbf{D}$ as defined earlier, the clustering problem requires estimating the assignment matrix $X$. In graph partitioning terms, this can be represented as
\begin{equation}
\label{eqn:pncx}
\begin{aligned}
\mathrm{maximize} \quad \epsilon(X) &= \frac{1}{K}\sum_{k=1}^K \frac{X_k^T \mathbf{A} X_k}{X_k^T \mathbf{D} X_k} \\
\mathrm{subject~to}\,\,\,\,\,\, \quad X &\in \{0,1\}^{N\times K}, \\
    X \boldsymbol{1}_K &= \boldsymbol{1}_N.
\end{aligned}
\end{equation}
Intuitively, the objective function $\epsilon(X)$ seeks to maximize the average ``link-ratio'', i.e., the fraction of all link weights in a group that stay within the group. The constraint $X \boldsymbol{1}_K = \boldsymbol{1}_N$ enforces the condition that each sample can belong to exactly 1 cluster. We will see later (cf. Section~\ref{sec:new}) how this constraint can be modified for our overlap-aware scenario.

The optimization problem in~\eqref{eqn:pncx} is NP-complete due to the discrete constraints on $X$. Instead of solving this original problem, we solve a relaxed version of this problem which ignores the constraints. Let 
\begin{equation}
\label{eqn:z_def}
Z = f(X) = X(X^T\mathbf{D}X)^{-\frac{1}{2}}. 
\end{equation}
It is easy to verify that $Z^T\mathbf{D}Z = I_K$. We can rewrite the above problem~\eqref{eqn:pncx}, by ignoring the constraints, as
\begin{equation}
\label{eqn:pncz}
\begin{aligned}
\mathrm{maximize}\,\,\,\,\,\,\, \quad \epsilon(Z) &= \text{tr}(Z^T\mathbf{A}Z) \\
\mathrm{subject~to} \quad Z^T\mathbf{D}Z &= I_K.
\end{aligned}
\end{equation}
Since Z has been relaxed into the continuous domain, the new optimization problem becomes tractable. Let 
\begin{equation}
P = \mathbf{D}^{-1}\mathbf{A},
\end{equation}
and suppose the eigen-decomposition of $P$ is given as $PV = VS$. Let $\Lambda^* = \mathrm{diag}(s_1,\ldots,s_K)$ and $Z^*$ contain the first $K$ columns of $V$. Then the global optimum of the problem described in equation~\eqref{eqn:pncz} occurs at
\begin{equation}
\label{eqn:z_sol}
    \{Z^*R: R^TR = I_K, PZ^* = Z^*\Lambda^*\}.
\end{equation}
This implies that the global optimum is not unique; rather, it is a subspace spanned by the first $K$ eigenvectors of $P$ through orthonormal matrices. 

The matrix $Z$ is a continuous solution to our clustering problem. To obtain a discrete solution, we solve for a discrete approximation for $Z$. First, we note from equation~\eqref{eqn:z_def} that
\begin{equation}
X = f^{-1}(Z) = \operatorname{Diag}\left({\mathrm{diag}}^{-\frac{1}{2}}\left(Z Z^{T}\right)\right) Z.
\end{equation}
Using this transformation, we can characterize the solution obtained in equation~\eqref{eqn:z_sol} as
\begin{equation}
     \{\Tilde{X}^*R: R^TR = I_K, \Tilde{X}^* = f^{-1}(Z^*)\}.
\end{equation}
Now, our discretization problem is to find an $X$ which approximates $\Tilde{X}^*R$ for some orthonormal $R$, such that $X$ obeys the discrete constraints from problem~\eqref{eqn:pncx}. Mathematically, this is formulated as
\begin{equation}
\label{eqn:pod}
\begin{aligned}
\mathrm{minimize} \quad \phi(X,R) &= \left\lVert X - \Tilde{X}^*R \right\rVert^2 \\
\mathrm{subject~to}\, \quad\quad\quad X &\in \{0,1\}^{N\times K}, \\
    X \boldsymbol{1}_K &= \boldsymbol{1}_N, \\
    R^TR &= I_K.
\end{aligned}
\end{equation}
It is difficult to minimize $\phi(X,R)$ jointly in $X$ and $R$, so we optimize it alternately in $X$ and $R$. Suppose we are given some $R^*$, then the problem~\eqref{eqn:pod} reduces to
\begin{equation}
\label{eqn:podx}
\begin{aligned}
\mathrm{minimize} \quad \phi(X) &= \left\lVert X - \Tilde{X}^*R^* \right\rVert^2 \\
\mathrm{subject~to} \quad\quad X &\in \{0,1\}^{N\times K}, \\
    X \boldsymbol{1}_K &= \boldsymbol{1}_N.
\end{aligned}
\end{equation}
The solution to this problem is given by non-maximal suppression, i.e.,
\begin{equation}
\label{eqn:podx_sol}
    X^{*}(i, l)=\left\langle l=\arg \max _{k \in[K]} \tilde{X}(i, k)\right\rangle, \quad i \in \{1,\ldots,N\}.
\end{equation}
Intuitively, we set the largest entry in each row as 1 and zero out all the others. This ensures that each sample belongs to exactly 1 cluster. In the next section, we will see how to reformulate problem~\eqref{eqn:podx} for the case when some samples can belong to more than one clusters.

Next, we fix $X^*$ and solve the following problem for $R^*$:
\begin{equation}
\label{eqn:podr}
\begin{aligned}
\mathrm{minimize} \quad \phi(R) &= \left\lVert X^* - \Tilde{X}^*R \right\rVert^2 \\
\mathrm{subject~to} \quad R^TR &= I_K.
\end{aligned}
\end{equation}
The solution to this problem is given by
\begin{equation}
    R^* = \Tilde{U}U^T,
\end{equation}
where $(U,\Omega,\tilde{U})$ is a singular value decomposition of $X^{*T}\tilde{X}^*$.

We solve the two problems~\eqref{eqn:podx} and \eqref{eqn:podr} iteratively until convergence, and finally return $X^*$ as the output of the clustering procedure.

\subsection{Extension to overlap-aware clustering}
\label{sec:new}

From equation~\eqref{eqn:overlap}, suppose the output of our overlap detector is given by $\mathbf{v}_{OL}$. Then, we can reformulate problem~\eqref{eqn:podx} to include the overlap constraint as
\begin{equation}
\label{eqn:podx_ol}
\begin{aligned}
\mathrm{minimize} \quad  \phi(X) &= \left\lVert X - \Tilde{X}^*R^* \right\rVert^2 \\
\mathrm{subject~to} \quad \quad X &\in \{0,1\}^{N\times K}, \\
    X \boldsymbol{1}_K &= \boldsymbol{1}_N + \mathbf{v}_{OL}.
\end{aligned}
\end{equation}

Intuitively, this solves the same optimum discretization problem, but the cluster exclusivity constraint has been modified to represent the condition that some samples may belong to more than one cluster. Similar to how we used non-maximal suppression in~\eqref{eqn:podx_sol} to solve the problem previously, the solution to the modified problem is again given by non-maximal suppression, with the exception that for samples belonging to more than one group, we set the largest two entries to 1, while zeroing out the others. Mathematically,
\begin{align}
\label{eqn:podx_newsol}
    \hat{X}^{*}(i, l)=X^{*}(i, l) + \mathbf{v}_{OL}^{(i)} \times \left\langle l=k^{\prime}_i\right\rangle, 
\end{align}
$~\forall i \in \{1,\ldots,N\}$, where $X^{*}(i, l)$ is defined in \eqref{eqn:podx_sol}, and $k^{\prime}_i$ is the index of the second largest element of $\Tilde{X}_i$. Although we only consider the case of 2-speaker overlaps here, it is easy to extend this method to the case of an arbitrary number of overlapping speakers.

\section{Overlap Detection}
\label{sec:overlap}

Our proposed overlap-aware diarization method relies heavily on the performance of $f(U)$, the overlap detector (OD). In this section, we detail an HMM-DNN based overlap detector. Our model is similar to the speech activity detector previously used in the CHiME-6 baseline system~\cite{Watanabe2020CHiME6CT}.

\begin{figure}[t]
\centering
\includegraphics[width=0.7\linewidth]{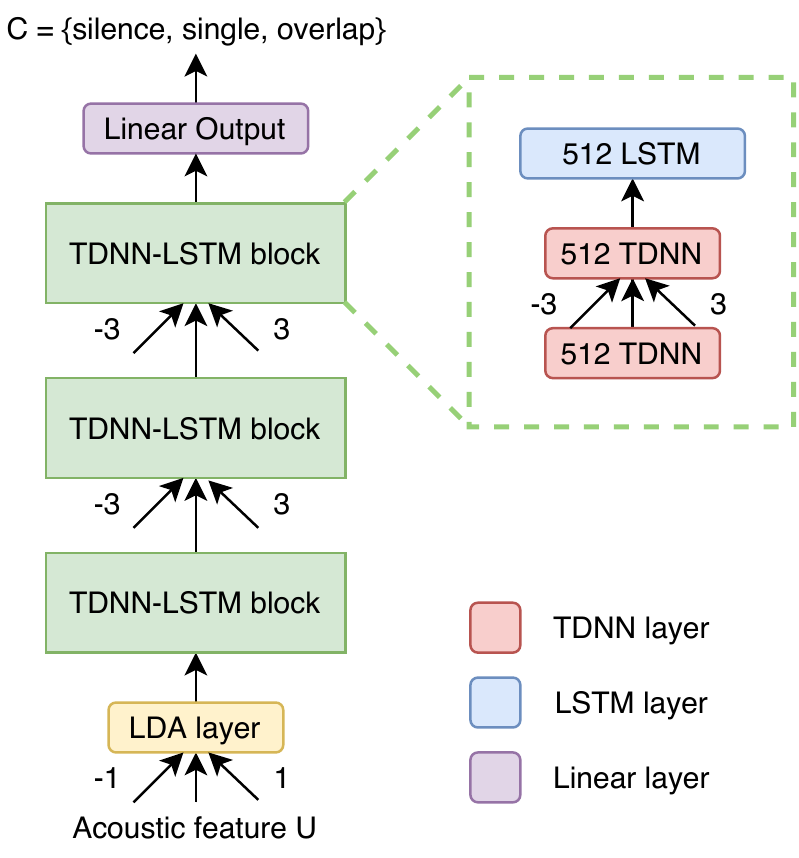}
\caption{Architecture of the neural network used for frame-level classification for overlap detection.}
\label{fig:od_nnet}
\vspace{-1em}
\end{figure}

We first trained a neural network classifier to assign each frame in an utterance a label from $\cal C$ = \{\textit{silence}, \textit{single}, \textit{overlap}\}, denoting silence, single speaker, or overlapping regions, respectively. We used the architecture shown in Figure~\ref{fig:od_nnet}, consisting of time-delay neural network (TDNN) layers to capture long temporal contexts \cite{Peddinti2015ATD}, interleaved with bidirectional long short term memory (BLSTM) layers with projection, to incorporate utterance-level statistics.

The posteriors obtained from the classifier were scaled with an external bias parameter tuned on the development data to reduce the false alarm rate. We then post-processed the per-frame classifier outputs to enforce minimum and maximum silence/single/overlap durations, by constructing a simple HMM whose state transition diagram encodes these constraints. Treating the per-frame posteriors like emission probabilities, we performed Viterbi decoding to obtain the most likely label-sequence. Furthermore, state transitions between the silence and overlap states were prohibited, mimicking real-world observations where it is highly unlikely for two speakers to start or stop speaking simultaneously. 

\section{Experimental Setup}
\label{sec:experiments}

\subsection{Datasets}

We performed experiments on two datasets -- the AMI meeting corpus~\cite{Carletta2005TheAM}, and the LibriCSS data~\cite{Chen2020ContinuousSS}. AMI consists of 100 hours of recorded meetings containing 4 speakers per session, with speech from close-talk, single distant microphone (SDM), and array microphones. For our experiments, we used the mix-headset recordings, which are obtained by summing the individual headset signals from the participants in the meeting. The dataset contains approximately 20\% overlap ratio, i.e., 20\% of the total speech contains overlaps. LibriCSS is a recently released corpus consisting of multi-channel audio recordings of ``simulated conversations.'' It comprises 10 sessions, where each session is approximately one hour long. Each session is made up of six 10-minute-long ``mini sessions'' that have different overlap ratios, ranging from 0 to 40\%, and contain 8 speakers. The recordings were made in a regular meeting room by using a seven-channel circular microphone array. For our experiments, we selected the recordings from the first channel of the array. We used this dataset to conduct a performance analysis of our proposed method on different overlap conditions. 

\vspace{-0.6em}
\subsection{Baselines}

We first have single-speaker baselines: (i) agglomerative hierarchical clustering (AHC) of x-vectors with probabilistic linear discriminant analysis (PLDA) scoring~\cite{Sell2018DiarizationIH}, (ii) spectral clustering of x-vectors with cosine scoring (using the Ng-Jordan-Weiss method)~\cite{Park2020AutoTuningSC}, and (iii) Bayesian HMM based x-vector clustering (VBx)~\cite{Diez2019BayesianHB,Dez2020OptimizingBH}. We used the same x-vector extractor for all the baselines (described in Section~\ref{sec:implementation}), such that the difference in their performance was only due to the clustering process. Furthermore, we used the same PLDA model (trained on a subset of the AMI training data) for the AHC and VBx baselines\footnote{The VBx diarization system has been shown to obtain significant gains with a PLDA interpolated between general data and in-domain data, but we did not use this method in this paper.}. For both these baselines, hyperparameters were tuned on the development set. No hyperparameter selection is required for spectral clustering since it is auto-tuned. We used a ground-truth VAD for these baselines as well as for our proposed method.

We also compare our approach with diarization methods that are not overlap-agnostic. These include: (i) overlap-aware VB resegmentation~\cite{Bullock2019OverlapawareDR} and (ii) region proposal networks (RPNs)~\cite{Huang2020SpeakerDW}. For the former, we present the official results from the paper, which uses a neural VAD. For RPNs, we filtered out non-speech regions using the ground truth VAD.

\vspace{-0.6em}
\subsection{Implementation details}
\label{sec:implementation}

\noindent
\textbf{X-vector extractor}. We used a similar x-vector extractor as described in earlier studies~\cite{GarciaRomero2017SpeakerDU,Park2020AutoTuningSC}. The model consists of TDNN layers with statistics pooling, and we extracted 128-dim embeddings from the pre-final layer. It was trained on VoxCeleb data~\cite{nagrani2017voxceleb} with simulated room impulse responses~\cite{ko2017study} using the Kaldi toolkit~\cite{povey2011kaldi}, and released as part of the CHiME-6 baseline~\cite{Watanabe2020CHiME6CT}.

\noindent
\textbf{Overlap detector}. We trained an HMM-DNN overlap detector (described in Section~\ref{sec:overlap}) using Kaldi. We used 40-dim MFCCs features as input, and trained the classifier on in-domain training data. For AMI, we used targets obtained from annotations of the official training set. Since LibriCSS does not have a corresponding training data, we generated simulated mixtures with reverberation using Librispeech training utterances~\cite{Panayotov2015LibrispeechAA} and used force-aligned targets for training our overlap detector. The decoding graph was created with additional constraints on the minimum (maximum) durations for single speaker and overlapping regions as 0.03s (10.0s) and 0.1s (5.0s), respectively. Note that our clustering method itself is independent of the overlap detector used. 

\noindent
\textbf{Overlap-aware spectral clustering}. We extended the spectral clustering algorithm in scikit-learn~\cite{scikit-learn} for our implementation. Since the overlap detector provides frame-level classification decisions whereas x-vectors were extracted for 1.5s segments, we assumed that a segment is ``overlapping'' if at least half of it lies in overlapping regions. For estimating the number of speakers $\widehat{K}$, we swept the binarization factor $p$ in the range from 2 to 20, similar to what was done in \cite{Park2020AutoTuningSC}.   

\vspace{-0.6em}
\section{Results and Discussion}
\label{sec:results}

\subsection{Overlap detection on AMI}

We present the results obtained by our overlap detector using 40-dim MFCC features on AMI mix-headset data in Table~\ref{tab:overlap_results}. We can see that the performance is comparable to previous studies on this dataset, without using waveform-level learned features. We used the output from this overlap detector for further experiments.

\begin{table}[t]
\centering
\caption{Overlap detection results on AMI mix-headset data, in terms of Precision (\%) and Recall (\%).}
\label{tab:overlap_results}
\begin{adjustbox}{max width=\linewidth}
\begin{tabular}{lcccc}
\toprule
\multicolumn{1}{c}{\multirow{2}{*}{\textbf{Model (feature type)}}} & \multicolumn{2}{c}{\textbf{Dev}} & \multicolumn{2}{c}{\textbf{Eval}} \\
\cmidrule(r{4pt}){2-3} \cmidrule(l){4-5}
\multicolumn{1}{c}{} & \textbf{Precision} & \textbf{Recall} & \textbf{Precision} & \textbf{Recall} \\
\midrule
ConvNet (Spectogram)~\cite{Kunesov2019DetectionOO} & 80.5 & 50.2 & 75.8 & 44.6 \\
E2E BLSTM (MFCC)~\cite{Bullock2019OverlapawareDR} & 90.0 & 52.5 & 91.9 & 48.4 \\
E2E BLSTM (SincNet)~\cite{Bullock2019OverlapawareDR} & 90.0 & 63.8 & 86.8 & 65.8 \\
\midrule
Our method (MFCC) & 83.9 & 68.5 & 86.4 & 65.2 \\
\bottomrule
\end{tabular}
\end{adjustbox}
\end{table}

\subsection{Diarization results for AMI}
\label{sec:ami_result}

Table~\ref{tab:diar_results} shows our proposed overlap-aware spectral clustering method compared with baselines, evaluated on the AMI mix-headset eval data. Using our overlap detector trained on the AMI train set, we were able to improve the DER from 28.3\% for the AHC/PLDA baseline, to 24.0\%, which is a relative improvement of 15.2\%. This compares favorably with the performance of other diarization methods like overlap-aware VB resegmentation and RPNs. Furthermore, it is possible to reduce the DER to 21.5\% using an oracle overlap detector.

\begin{table}[t]
\centering
\caption{Diarization results on AMI mixed-headset eval set, in terms of missed speech (MS), false alarm (FA), speaker confusion (Conf.), and diarization error rate (DER). For all the recordings in the test set, the NME-based speaker counting approach estimated between 3 and 6 speakers, which is close to the oracle count of 4 speakers.}
\label{tab:diar_results}
\begin{adjustbox}{max width=\linewidth}
\begin{tabular}{@{}lcccc@{}}
\toprule
\multicolumn{1}{c}{\textbf{Method}} & \textbf{MS} & \textbf{FA} & \textbf{Conf.} & \textbf{DER} \\ \midrule
AHC/PLDA & 19.9 & 0.0 & 8.4 & 28.3 \\
Spectral/cosine & 19.9 & 0.0 & 7.0 & 26.9 \\
VBx~\cite{Diez2019BayesianHB} & 19.9 & 0.0 & 6.3 & 26.2 \\ \midrule
VB resegmentation~\cite{Bullock2019OverlapawareDR} & 13.0 & 3.6 & 7.2 & 23.8 \\
RPN~\cite{Huang2020SpeakerDW} & 9.5 & 7.7 & 8.3 & 25.5 \\ \midrule
Our method & 11.3 & 2.2 & 10.5 & 24.0 \\
Our method + oracle OD & 7.4 & 1.8 & 12.3 & \underline{21.5} \\ 
Our method + noise aug. & 11.3 & 2.2 & 10.1 & \textbf{23.6} \\ \bottomrule
\end{tabular}
\end{adjustbox}
\vspace{-1em}
\end{table}

\begin{figure}[t]
\begin{subfigure}{0.49\linewidth}
\centering
\includegraphics[width=\linewidth]{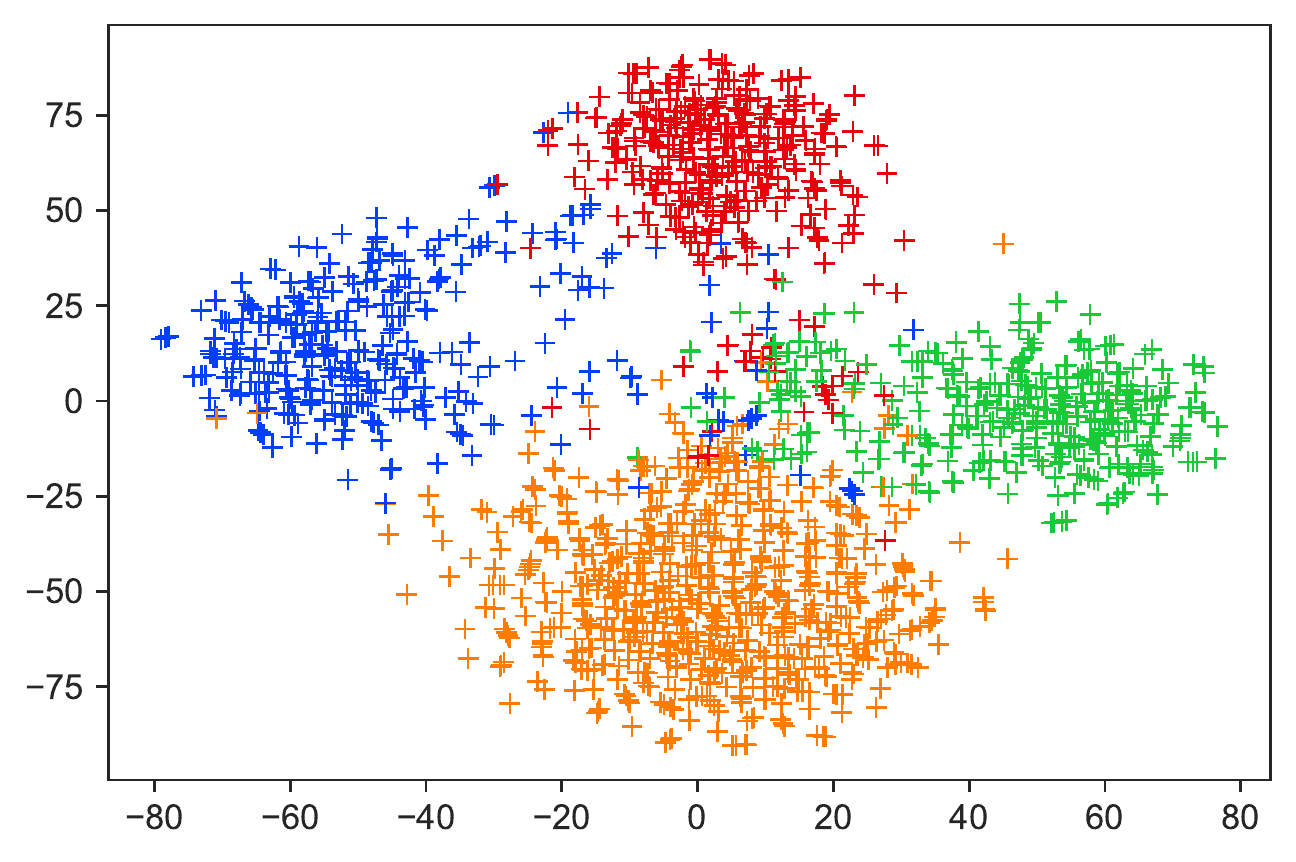}
\caption{}
\label{fig:non_ovl}
\end{subfigure}
\begin{subfigure}{0.49\linewidth}
\centering
\includegraphics[width=\linewidth]{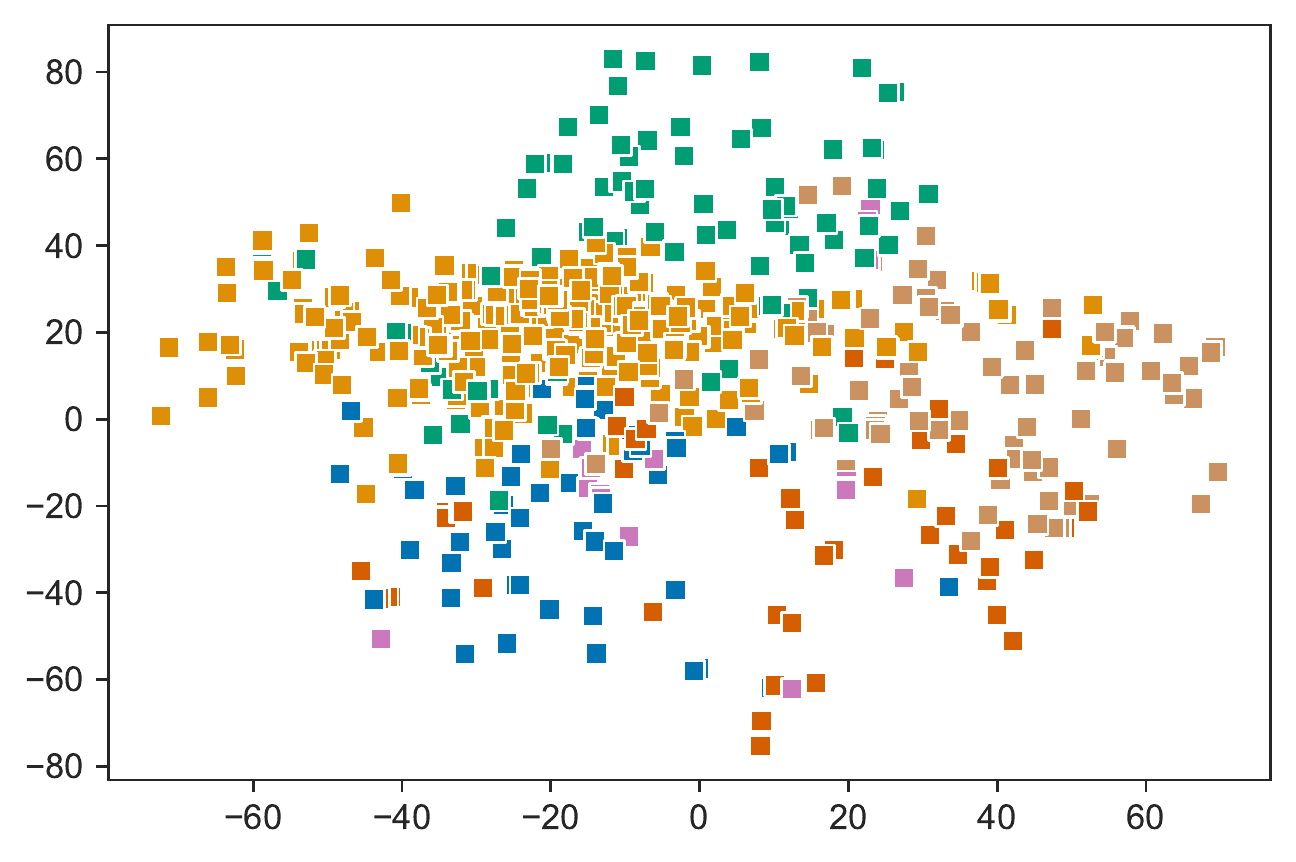}
\caption{}
\label{fig:ovl}
\end{subfigure}\hfill
\caption{T-SNE plots of x-vector embeddings for (a) non-overlapping, and (b) overlapping segments for the recording \texttt{EN2002a} in the AMI eval set (containing 4 speakers). Colors denote the speaker assigned to the segment. For (b), each color represents a distinct pair of speaker labels, resulting in 6 differently colored clusters.}
\label{fig:xvec_plot}
\end{figure}

A detailed analysis of the results reveals that although the missed speech reduces substantially (from 19.9\% to 11.3\%) as a result of overlap detection, there is also a significant increase in speaker confusion errors (from 8.4\% to 10.5\%). We conjecture that since the x-vector extractor was trained only on single-speaker utterances, a mismatch in the overlap regions of the recording results in noisy samples. The speaker confusion improved by 0.4\% when we used an x-vector extractor trained with noise augmentation, using noises from the MUSAN corpus~\cite{Snyder2015MUSANAM}. To verify our hypothesis further, we show the T-SNE plots for the non-overlapping and overlapping segments in Fig.~\ref{fig:non_ovl} and \ref{fig:ovl}, respectively. We can see that while the embeddings for the non-overlapping segments are well separated, those for overlapping segments may often be noisy, leading to clustering errors.

\subsection{Analysis on LibriCSS}

Table~\ref{tab:libricss} shows a breakdown of diarization errors obtained by our system, compared with some of the baselines. It is evident that as the overlap ratio increases from 0 to 40\%, the difference in performance becomes more significant. On average, our method provided a 42.9\% relative DER improvement compared to a baseline AHC system, and this increased to 46.0\% relative on using an oracle overlap detector. We note here that since LibriCSS does not have a corresponding training set, the PLDA was trained on Librispeech utterances. The mismatch between clean training data versus overlapping mixtures at test time may be particularly detrimental to the performance of the AHC system. As the overlap ratio increases, RPN performs better than our method. We again attribute this to the fact that our RPN model is trained on closely matched overlapping speech, whereas the x-vector extractor was trained on single-speaker utterances, which results in a higher speaker confusion (cf. Section~\ref{sec:ami_result}).

\begin{table}[t]
\centering
\caption{Diarization performance on LibriCSS evaluation set (sessions 2-10), evaluated condition-wise, in terms of \% DER. 0S and 0L refer to 0\% overlap with short and long inter-utterance silences, respectively. Our overlap detector obtained 96.3\% precision and 83.8\% recall on this data. RPN does not use ground truth VAD.}
\label{tab:libricss}
\begin{adjustbox}{max width=\linewidth}
\begin{tabular}{@{}lccccccc@{}}
\toprule
\multicolumn{1}{c}{\multirow{2}{*}{\textbf{Method}}} & \multicolumn{6}{c}{\textbf{Overlap ratio in \%}} & \multicolumn{1}{c}{\multirow{2}{*}{\textbf{Average}}} \\
\cmidrule(r{4pt}){2-7}
\multicolumn{1}{c}{} & \multicolumn{1}{c}{\textbf{0L}} & \multicolumn{1}{c}{\textbf{0S}} & \multicolumn{1}{c}{\textbf{10}} & \multicolumn{1}{c}{\textbf{20}} & \multicolumn{1}{c}{\textbf{30}} & \multicolumn{1}{c}{\textbf{40}} & \multicolumn{1}{c}{} \\
\midrule
AHC/PLDA & 5.1 & 2.7 & 11.5 & 19.3 & 22.7 & 28.8 & 16.3 \\
Spectral/cosine & 2.2 & 1.9 & 8.5 & 13.5 & 19.0 & 23.4 & 12.6 \\
RPN & 4.5 & 9.1 & 8.3 & 6.7 & 11.6 & 14.2 & 9.5 \\
\midrule
Our method & 2.6 & 3.4 & 6.8 & 10.0 & 13.9 & 15.2 & 9.3 \\
 + oracle OD & 2.2 & 3.3 & 6.7 & 9.6 & 12.9 & 14.4 & 8.8 \\
\bottomrule
\end{tabular}
\end{adjustbox}
\vspace{-1em}
\end{table}

\vspace{-0.6em}
\section{Related Work}
\label{sec:related}


Spectral clustering was first applied to speaker diarization in~\cite{Ning2006ASC} using the Ng-Jordan-Weiss (NJW) algorithm~\cite{Ng2001OnSC}. \cite{Bassiou2010SpeakerDE} extended this to the case of an unknown number of speakers by using the eigengap criterion. Agglomerative and spectral clustering methods for meeting diarization were compared in~\cite{Luque2012OnTU}. After i-vectors were proposed for speaker recognition~\cite{Dehak2011FrontEndFA}, they were combined with cosine scoring and spectral clustering to perform diarization in~\cite{Shum2012OnTU}. 
It was further observed that spectral clustering was more robust to non-stationary environmental noise compared to other clustering methods~\cite{Tawara2015ACS}. 
More recently, with the ubiquitousness of deep neural networks, several researchers have proposed methods to incorporate DNNs with spectral clustering. \cite{Lin2019LSTMBS} proposed a supervised method to measure the similarity matrix between all segments of an audio recording with BLSTMs, and applied spectral clustering on top of the similarity matrix. Other approaches use DNN-based speaker embeddings, such as x-vectors~\cite{Snyder2018XVectorsRD}, to compute the similarty matrix between segment pairs~\cite{Park2020AutoTuningSC,Medennikov2020TargetSpeakerVA}. Additionally, \cite{Park2020AutoTuningSC} introduced $p$-binarization and normalized maximum eigengap (NME) techniques to automatically estimate the number of speakers in the recording.

Speaker diarization in overlapping settings has been studied extensively (cf. Section~\ref{sec:intro}). However, to the best of our knowledge, there is no prior work on incorporating overlap awareness into spectral clustering based diarization. The recently proposed target speaker voice activity detection (TS-VAD)~\cite{Medennikov2020TargetSpeakerVA} method uses x-vector based spectral clustering for initial estimate of speaker i-vectors, and thereafter performs frame-level multi-label classification to predict speaker activities in a speech frame. 

\vspace{-0.6em}
\section{Conclusion}
\label{sec:conclusion}

We proposed a new method for overlap-aware speaker diarization using spectral clustering. We leveraged an external overlap detector to identify the overlapping subsegments, and then assigned these segments to multiple speakers during clustering. The clustering approach itself was reformulated by first relaxing the discrete constraints, and then solving an optimal discretization problem with the additional overlap constraints. Our method provided significant improvements over conventional single-speaker clustering models on the AMI meeting corpus, and was competitive with other overlap-aware diarization methods. Analysis on LibriCSS showed that overlapping regions benefit strongly from this approach, although speaker confusion may increase due to an inadequate speaker embedding extractor. We conjecture that this problem may be alleviated by training the x-vector extractor additionally on overlapping segments. This investigation is left as future work.

\vspace{-0.6em}
\small
\section{Acknowledgment}

The authors thank Takuya Yoshioka for providing simulation scripts for the LibriCSS training data, and Leibny Paola Garc\'ia-Perera for helpful discussions and insights. This work was partially supported by research grants from Johns Hopkins Applied Physics Laboratory, Government of Israel, Hitachi Ltd., Japan, and Nanyang Technological University, Singapore.

\bibliographystyle{IEEEbib}
\bibliography{mybib}

\begin{thebibliography}{10}

\bibitem{Mir2012SpeakerDA}
Xavier~Anguera Mir{\'o}, Simon Bozonnet, Nicholas W.~D. Evans, Corinne
  Fredouille, Gerald Friedland, and Oriol Vinyals,
\newblock ``Speaker diarization: A review of recent research,''
\newblock {\em IEEE Transactions on Audio, Speech, and Language Processing},
  vol. 20, pp. 356--370, 2012.

\bibitem{Tranter2006AnOO}
Sue Tranter and Douglas~A. Reynolds,
\newblock ``An overview of automatic speaker diarization systems,''
\newblock {\em IEEE Transactions on Audio, Speech, and Language Processing},
  vol. 14, pp. 1557--1565, 2006.

\bibitem{GarciaRomero2017SpeakerDU}
Daniel Garcia-Romero, David Snyder, Gregory Sell, Daniel Povey, and Alan
  McCree,
\newblock ``Speaker diarization using deep neural network embeddings,''
\newblock {\em 2017 IEEE International Conference on Acoustics, Speech and
  Signal Processing (ICASSP)}, pp. 4930--4934, 2017.

\bibitem{Sun2018SpeakerDW}
Lei Sun, Jun Du, Chao Jiang, Xueyang Zhang, Shan He, Bing Yin, and Chin-Hui
  Lee,
\newblock ``Speaker diarization with enhancing speech for the first dihard
  challenge,''
\newblock in {\em INTERSPEECH}, 2018.

\bibitem{Dehak2011FrontEndFA}
Najim Dehak, Patrick Kenny, R{\'e}da Dehak, Pierre Dumouchel, and Pierre
  Ouellet,
\newblock ``Front-end factor analysis for speaker verification,''
\newblock {\em IEEE Transactions on Audio, Speech, and Language Processing},
  vol. 19, pp. 788--798, 2011.

\bibitem{Variani2014DeepNN}
Ehsan Variani, Xin Lei, Erik McDermott, Ignacio Lopez-Moreno, and Javier
  Gonzalez-Dominguez,
\newblock ``Deep neural networks for small footprint text-dependent speaker
  verification,''
\newblock {\em ICASSP}, pp. 4052--4056, 2014.

\bibitem{Snyder2018XVectorsRD}
David Snyder, Daniel Garcia-Romero, Gregory Sell, Daniel Povey, and Sanjeev
  Khudanpur,
\newblock ``X-vectors: Robust {DNN} embeddings for speaker recognition,''
\newblock {\em 2018 IEEE International Conference on Acoustics, Speech and
  Signal Processing (ICASSP)}, pp. 5329--5333, 2018.

\bibitem{Sell2015DiarizationRI}
Gregory Sell and Daniel Garcia-Romero,
\newblock ``Diarization resegmentation in the factor analysis subspace,''
\newblock {\em 2015 IEEE International Conference on Acoustics, Speech and
  Signal Processing (ICASSP)}, pp. 4794--4798, 2015.

\bibitem{boakye2008overlapped}
Kofi Boakye, Beatriz Trueba-Hornero, Oriol Vinyals, and Gerald Friedland,
\newblock ``Overlapped speech detection for improved speaker diarization in
  multiparty meetings,''
\newblock in {\em 2008 IEEE International Conference on Acoustics, Speech and
  Signal Processing}. IEEE, 2008, pp. 4353--4356.

\bibitem{Huijbregts2009SpeechOD}
Marijn Huijbregts, David~A. van Leeuwen, and Franciska de~Jong,
\newblock ``Speech overlap detection in a two-pass speaker diarization
  system,''
\newblock in {\em INTERSPEECH}, 2009.

\bibitem{Yella2012SpeakerDO}
Sree~Harsha Yella and Fabio Valente,
\newblock ``Speaker diarization of overlapping speech based on silence
  distribution in meeting recordings,''
\newblock in {\em INTERSPEECH}, 2012.

\bibitem{Geiger2013DetectingOS}
J{\"u}rgen~T. Geiger, Florian Eyben, Bj{\"o}rn~W. Schuller, and Gerhard Rigoll,
\newblock ``Detecting overlapping speech with long short-term memory recurrent
  neural networks,''
\newblock in {\em INTERSPEECH}, 2013.

\bibitem{Andrei2017DetectingOS}
Valentin Andrei, Horia Cucu, and Corneliu Burileanu,
\newblock ``Detecting overlapped speech on short timeframes using deep
  learning,''
\newblock in {\em INTERSPEECH}, 2017.

\bibitem{Hagerer2017EnhancingLR}
Gerhard Hagerer, Vedhas Pandit, Florian Eyben, and Bj{\"o}rn~W. Schuller,
\newblock ``Enhancing {LSTM} {RNN}-based speech overlap detection by
  artificially mixed data,''
\newblock in {\em Semantic Audio}, 2017.

\bibitem{Kunesov2019DetectionOO}
Marie Kunesov{\'a}, Marek Hr{\'u}z, Zbynek Zaj{\'i}c, and Vlasta Radov{\'a},
\newblock ``Detection of overlapping speech for the purposes of speaker
  diarization,''
\newblock in {\em SPECOM}, 2019.

\bibitem{Bullock2019OverlapawareDR}
Latan{\'e} Bullock, Herv{\'e} Bredin, and L.~Paola Garc{\'i}a-Perera,
\newblock ``Overlap-aware diarization: resegmentation using neural end-to-end
  overlapped speech detection,''
\newblock {\em ArXiv}, vol. abs/1910.11646, 2019.

\bibitem{Dez2018SpeakerDB}
Mireia D{\'i}ez, Luk{\'a}s Burget, and Pavel Matejka,
\newblock ``Speaker diarization based on {Bayesian} {HMM} with eigenvoice
  priors,''
\newblock in {\em Odyssey}, 2018.

\bibitem{Fujita2020EndtoEndND}
Yusuke Fujita, Shinji Watanabe, Shota Horiguchi, Yawen Xue, and Kenji
  Nagamatsu,
\newblock ``End-to-end neural diarization: Reformulating speaker diarization as
  simple multi-label classification,''
\newblock {\em ArXiv}, vol. abs/2003.02966, 2020.

\bibitem{Huang2020SpeakerDW}
Zili Huang, Shinji Watanabe, Yusuke Fujita, Paola Garc{\'i}a, Yiwen Shao,
  Daniel Povey, and Sanjeev Khudanpur,
\newblock ``Speaker diarization with region proposal network,''
\newblock {\em ArXiv}, vol. abs/2002.06220, 2020.

\bibitem{Yu2003MulticlassSC}
Stella~X. Yu and Jianbo Shi,
\newblock ``Multiclass spectral clustering,''
\newblock {\em Proceedings Ninth IEEE International Conference on Computer
  Vision}, pp. 313--319 vol.1, 2003.

\bibitem{Park2020AutoTuningSC}
Tae~Jin Park, Kyu~J. Han, Manoj Kumar, and Shrikanth~S. Narayanan,
\newblock ``Auto-tuning spectral clustering for speaker diarization using
  normalized maximum eigengap,''
\newblock {\em IEEE Signal Processing Letters}, vol. 27, pp. 381--385, 2020.

\bibitem{Luxburg2007ATO}
Ulrike von Luxburg,
\newblock ``A tutorial on spectral clustering,''
\newblock {\em Statistics and Computing}, vol. 17, pp. 395--416, 2007.

\bibitem{Stewart1990MatrixPT}
G.~W. Stewart and Ji-Guang Sun,
\newblock ``Matrix perturbation theory,''
\newblock 1990.

\bibitem{Fiedler1973AlgebraicCO}
Miroslav Fiedler,
\newblock ``Algebraic connectivity of graphs,''
\newblock 1973.

\bibitem{Ng2001OnSC}
Andrew~Y. Ng, Michael~I. Jordan, and Yair Weiss,
\newblock ``On spectral clustering: Analysis and an algorithm,''
\newblock in {\em NIPS}, 2001.

\bibitem{Medennikov2020TargetSpeakerVA}
Ivan Medennikov, Maxim Korenevsky, Tatiana Prisyach, Yuri~Y. Khokhlov, Mariya
  Korenevskaya, Ivan Sorokin, Tatiana~V. Timofeeva, Anton Mitrofanov, Andrei
  Andrusenko, Ivan Podluzhny, Aleksandr Laptev, and Aleksei Romanenko,
\newblock ``Target-speaker voice activity detection: a novel approach for
  multi-speaker diarization in a dinner party scenario,''
\newblock {\em ArXiv}, vol. abs/2005.07272, 2020.

\bibitem{Watanabe2020CHiME6CT}
Shinji Watanabe, Michael Mandel, Jon Barker, and Emmanuel Vincent,
\newblock ``Chime-6 challenge: Tackling multispeaker speech recognition for
  unsegmented recordings,''
\newblock {\em ArXiv}, vol. abs/2004.09249, 2020.

\bibitem{Peddinti2015ATD}
Vijayaditya Peddinti, Daniel Povey, and Sanjeev Khudanpur,
\newblock ``A time delay neural network architecture for efficient modeling of
  long temporal contexts,''
\newblock in {\em INTERSPEECH}, 2015.

\bibitem{Carletta2005TheAM}
Jean Carletta, Simone Ashby, Sebastien Bourban, Mike Flynn, Ma{\"e}l Guillemot,
  Thomas Hain, Jaroslav Kadlec, Vasilis Karaiskos, Wessel Kraaij, Melissa
  Kronenthal, Guillaume Lathoud, Mike Lincoln, Agnes~Lisowska Masson, Iain
  McCowan, Wilfried Post, Dennis Reidsma, and Pierre Wellner,
\newblock ``The ami meeting corpus: A pre-announcement,''
\newblock in {\em MLMI}, 2005.

\bibitem{Chen2020ContinuousSS}
Zhuo Chen, Takuya Yoshioka, Liang Lu, Tianyan Zhou, Zhong Meng, Yi~Luo, J.~Wu,
  and Jinyu Li,
\newblock ``Continuous speech separation: Dataset and analysis,''
\newblock {\em ICASSP 2020 - 2020 IEEE International Conference on Acoustics,
  Speech and Signal Processing (ICASSP)}, pp. 7284--7288, 2020.

\bibitem{Sell2018DiarizationIH}
Gregory Sell, David Snyder, Alan McCree, Daniel Garcia-Romero, Jes{\'u}s
  Villalba, Matthew Maciejewski, Vimal Manohar, Najim Dehak, Daniel Povey,
  Shinji Watanabe, and Sanjeev Khudanpur,
\newblock ``Diarization is hard: Some experiences and lessons learned for the
  jhu team in the inaugural dihard challenge,''
\newblock in {\em INTERSPEECH}, 2018.

\bibitem{Diez2019BayesianHB}
Mireia Diez, Luk{\'a}s Burget, Shuai Wang, Johan Rohdin, and Jan Cernock{\'y},
\newblock ``Bayesian hmm based x-vector clustering for speaker diarization,''
\newblock in {\em INTERSPEECH}, 2019.

\bibitem{Dez2020OptimizingBH}
Mireia D{\'i}ez, Luk{\'a}s Burget, Federico Landini, Shuai Wang, and Jan
  Cernock{\'y},
\newblock ``Optimizing bayesian hmm based x-vector clustering for the second
  dihard speech diarization challenge,''
\newblock {\em ICASSP 2020 - 2020 IEEE International Conference on Acoustics,
  Speech and Signal Processing (ICASSP)}, pp. 6519--6523, 2020.

\bibitem{nagrani2017voxceleb}
Arsha Nagrani, Joon~Son Chung, and Andrew Zisserman,
\newblock ``{VoxCeleb}: a large-scale speaker identification dataset,''
\newblock {\em arXiv preprint arXiv:1706.08612}, 2017.

\bibitem{ko2017study}
Tom Ko, Vijayaditya Peddinti, Daniel Povey, Michael~L Seltzer, and Sanjeev
  Khudanpur,
\newblock ``A study on data augmentation of reverberant speech for robust
  speech recognition,''
\newblock in {\em ICASSP}. IEEE, 2017, pp. 5220--5224.

\bibitem{povey2011kaldi}
Daniel Povey, Arnab Ghoshal, Gilles Boulianne, Lukas Burget, Ondrej Glembek,
  Nagendra Goel, Mirko Hannemann, Petr Motlicek, Yanmin Qian, Petr Schwarz,
  et~al.,
\newblock ``The {Kaldi} speech recognition toolkit,''
\newblock in {\em IEEE 2011 workshop on automatic speech recognition and
  understanding}. IEEE Signal Processing Society, 2011, number CONF.

\bibitem{Panayotov2015LibrispeechAA}
Vassil Panayotov, Guoguo Chen, Daniel Povey, and Sanjeev Khudanpur,
\newblock ``Librispeech: An asr corpus based on public domain audio books,''
\newblock {\em 2015 IEEE International Conference on Acoustics, Speech and
  Signal Processing (ICASSP)}, pp. 5206--5210, 2015.

\bibitem{scikit-learn}
F.~Pedregosa, G.~Varoquaux, A.~Gramfort, V.~Michel, B.~Thirion, O.~Grisel,
  M.~Blondel, P.~Prettenhofer, R.~Weiss, V.~Dubourg, J.~Vanderplas, A.~Passos,
  D.~Cournapeau, M.~Brucher, M.~Perrot, and E.~Duchesnay,
\newblock ``Scikit-learn: Machine learning in {P}ython,''
\newblock {\em Journal of Machine Learning Research}, vol. 12, pp. 2825--2830,
  2011.

\bibitem{Snyder2015MUSANAM}
David Snyder, Guoguo Chen, and Daniel Povey,
\newblock ``Musan: A music, speech, and noise corpus,''
\newblock {\em ArXiv}, vol. abs/1510.08484, 2015.

\bibitem{Ning2006ASC}
Huazhong Ning, Ming Liu, Hao Tang, and Thomas~S. Huang,
\newblock ``A spectral clustering approach to speaker diarization,''
\newblock in {\em INTERSPEECH}, 2006.

\bibitem{Bassiou2010SpeakerDE}
Nikoletta Bassiou, Vassiliki Moschou, and Constantine Kotropoulos,
\newblock ``Speaker diarization exploiting the eigengap criterion and cluster
  ensembles,''
\newblock {\em IEEE Transactions on Audio, Speech, and Language Processing},
  vol. 18, pp. 2134--2144, 2010.

\bibitem{Luque2012OnTU}
Jordi Luque and Javier Hernando,
\newblock ``On the use of agglomerative and spectral clustering in speaker
  diarization of meetings,''
\newblock in {\em Odyssey}, 2012.

\bibitem{Shum2012OnTU}
Stephen Shum, Najim Dehak, and Jim Glass,
\newblock ``On the use of spectral and iterative methods for speaker
  diarization,''
\newblock in {\em INTERSPEECH}, 2012.

\bibitem{Tawara2015ACS}
Naohiro Tawara, Tetsuji Ogawa, and Tetsunori Kobayashi,
\newblock ``A comparative study of spectral clustering for i-vector-based
  speaker clustering under noisy conditions,''
\newblock {\em 2015 IEEE International Conference on Acoustics, Speech and
  Signal Processing (ICASSP)}, pp. 2041--2045, 2015.

\bibitem{Lin2019LSTMBS}
Qingjian Lin, Ruiqing Yin, Ming Li, Herv{\'e} Bredin, and Claude Barras,
\newblock ``Lstm based similarity measurement with spectral clustering for
  speaker diarization,''
\newblock in {\em INTERSPEECH}, 2019.

\end{thebibliography}

\end{document}